\documentclass[conference,a4paper]{IEEEtran}

\usepackage[utf8]{inputenc} 
\usepackage[T1]{fontenc}
\usepackage{url}              
\usepackage{cite}             

\usepackage[cmex10]{amsmath}  
\usepackage{mleftright}       
\mleftright                   

\usepackage{graphicx}         

\usepackage{booktabs}         

\usepackage{algorithm}    

\usepackage{xcolor}
\usepackage[braket, qm]{qcircuit}
\usepackage{dsfont}
\usepackage{diagbox}
\usepackage{arydshln}

\begin{document}

\title{Performance of Uncoded Implementation of Grover's Algorithm on Today's Quantum Processors
} 

 \author{
   \IEEEauthorblockN{Yunos El~Kaderi$^{*\dagger}$ and Andreas Honecker$^*$ and Iryna Andriyanova$^{\dagger}$}
   \IEEEauthorblockA{$^*$LPTM UMR8089 Laboratory and $^{\dagger}$ETIS UMR8051 Laboratory\\
                     CY Cergy Paris University, ENSEA, CNRS\\
                     Cergy, France                     }
}

\maketitle

\begin{abstract}
This work tests the performance of Grover's search circuits on some IBM superconducting quantum devices in case of the size of search space $N=2^4$ and $N=2^5$.   
Ideally, we expect to get an outcome probability distribution that is clearly peaked at the goal (marked) state. However, the quantum circuit executed on real devices is vulnerable to noise which leads to fluctuations in the results. 
The contributions of the paper are therefore the following: a) it presents two new Grover's search circuits for $N=16$ which were not yet reported in the state of the art; b) it shows performance difference between simulation results and results obtained on real devices; c) it shows the need of adding error-correction on the circuit for $N\ge 2^5$. 
\end{abstract}

\section{Introduction}
\label{sec:introduction}
Quantum computing has become one of the hottest topics of the last decade in physics, mathematics and computer science fields. This comes from the deployment of Noisy Intermediate Scale Quantum (NISQ) devices, that allow to accelerate execution of a number of algorithms. 
In contrast to Large Scale Quantum (LSQ) systems, NISQ devices are not based on fault-tolerant quantum circuits that use error-correction codes, but make use of error mitigation techniques. 

It is agreed that quantum error correction will play an essential role in development of LSQ systems. A lot of literature is devoted to the topic, and the research field is quickly developing in last years \cite{Girvin2021-lectures,eczoo_quantum}. It would be also interesting to investigate whether error-control coding might be useful for NISQ technologies. To the best of our knowledge, no tests with the use of error correction have been performed on NISQ devices until very recently. At the end of 2022, the authors of \cite{Pokharel2022} have reported a performance improvement while using a [[4,2,2]] error-detection code.  

The goal of this work is to show performance limitations of quantum algorithms implemented on available NISQ devices, and to discuss their improvements, possibly with the use of error correction, in line with recent results from \cite{Pokharel2022}. 
For better understanding, we are going to focus on a particular case of the Grover's search algorithm (GSA). This algorithm has been suggested in \cite{Grover96} to search for a marked element over an unsorted dataset and theoretically outperforms classical search algorithms (the GSA has complexity growing as $\sqrt{N}$ with the dataset size $N$, while the best classical search algorithm behaves as $O(N)$). An information-theoretic analysis of the GSA can be found in \cite{Arikan03}.

The reason of choosing the GSA is two-fold. Firstly, it makes part of the most common algorithms used for tests on NISQ devices since the last two years, and a number of references is available for a benchmark. Secondly, the Grover's algorithm is the one with the largest depth of quantum circuit, therefore it is the most sensitive to noise \cite{Lubinski2021}.  
In fact, there exist recent publications presenting results on real devices where the noise overshadows the signal. This observation will be discussed later in the paper.

The paper is organized as follows. 
Section \ref{sec:grover} presents the Grover's search algorithm, one circuit implementation from the state of the art and two circuits suggested by the authors of the paper. 
Section \ref{sec:results} presents numerical results obtained by running the circuits over NISQ devices and Section \ref{sec:discussion} concludes the paper. 
It is to note that the circuits have been first run on the Qiskit simulator \cite{Qiskit} and then on IBM quantum devices of $7$ and $27$ qubits.

\subsection{State of the art}
\label{sec:state-art}

\textit{NISQ implementations.} Among a large amount of results on the NISQ implementation, appeared in the last several years, let us report the most relevant tests of the Grover's algorithm. 
A 4-qubit String Detection Problem (SDP) circuit has been introduced and tested in \cite{Wang2022}, for a multi-target search, both on simulator and on quantum devices. It is to mention the GSA with $N=8$, reported there, first failed to find correct marked states of the system when carried on quantum devices, but it further succeeded with a slight modification. 
In order to make the GSA with $N=8$ to succeed in searching for a correct marked state, a circuit with auxiliary qubits has been presented in \cite{Vlasic2022}.  This introduced a GSA with a bigger number of states than $N$, where the redundant states are further removed by amplitude suppression. 
Furthermore, a GSA with $N=8$ and one single ancilla qubit has been implemented on IBM quantum devices in
\cite{3qubitGrover-Roy-2020}, and a GSA with $N=8$ both with and without ancilla qubits has been executed on IonQ\footnote{Based on trapped atomic ion qubits} quantum devices  
in \cite{3qubitGrover-IonQ-Figgatt-2017}. 

\textit{Error correction.}
Regarding the application of error-correction techniques to the GSA implementation, the use of an error-correction code was considered in \cite{Botsinis2016}, where the authors have applied two quantum BCH codes of parameters [[7,1]] and [[15,7]] in order to correct errors after each stage of the GSA with $N=128$ and $N=1024$. In \cite{Botsinis2016}, the performance evaluation of a noisy GSA has been done by simulation, assuming a depolarizing channel of probability $p$ within any two GSA stages, and no quantum devices have been used at that time.     
Very recently, a NISQ implementation of the GSA with $N=4$ using a [[4,2,2]] error-detection code has been reported in the literature \cite{Pokharel2022}. 

\subsection{Useful notation}
\label{sec:notation}
Let us introduce some notation which will be used though the paper.
Denote two pure states by $\ket{0}=\begin{pmatrix}
        1\\
        0
    \end{pmatrix}$ and $\ket{1} =
    \begin{pmatrix}
        0\\
        1
    \end{pmatrix}$, and a superposed state by $\ket{q} = a\ket{0}+b\ket{1}$, with $|a|^2+|b|^2=1$. 
One represents     
a system of $n$ qubits as $\ket{\textsl{system}} = \ket{q_1}\otimes\ket{q_2}\otimes...\ket{q_N}$, where $\otimes$ is the Kronecker product.
Thus, if $\ket{q_1} = \ket{q_2} = ... = \ket{q_n} = \ket{q}$,  then $\ket{\textsl{system}} = \ket{q}^{\otimes n}$.
Finally, let $\ket{+} = \frac{\ket{0}+\ket{1}}{\sqrt{2}}$ and $\ket{-} = \frac{\ket{0}-\ket{1}}{\sqrt{2}}$.

Let us denote by $H$, $X$ and $Z$ the gates given by respective unitary matrices $$\frac{1}{\sqrt{2}}\begin{pmatrix}
        1&1\\
        -1&1
    \end{pmatrix}, \quad
   \begin{pmatrix}
        0&1\\
        1&0
    \end{pmatrix}, \textrm{ and }
    \begin{pmatrix}
        1&0\\
        0&-1
    \end{pmatrix}.$$
Also, the $CX$ the operator is given by the $4\times 4$ unitary matrix
$$CX=I \oplus X,$$
where $I$ is the $2\times 2$ identity matrix and $\oplus$ is the direct sum.
Similarly, let $CZ=I\oplus Z$, and more generally let $C^iX= \underbrace{I\oplus \ldots \oplus I}_\text{$i$ times} \oplus X$ and $C^iZ= \underbrace{I\oplus \ldots \oplus I}_\text{$i$ times} \oplus Z$ for all $i\ge 1$.     

\section{Grover's Algorithm: theory and implementation}
\label{sec:grover}

\subsection{GSA in Brief}
\label{sec:algo}
The GSA can be briefly described as follows.
\begin{algorithm}\label{GSA}
\caption{\label{alg:Grover}Grover's search algorithm with $N=2^n$ \cite{Grover96}}

{\bf Initialization:}
\begin{itemize}
\item Compute
$R = \lfloor\pi\sqrt{N}/2 - 1/2\rfloor$.
\item Prepare two states of $n$ qubits each:\\
$\ket{s_{\textsl{initial}}}=\ket{0}^{\otimes n}$ \\
$\ket{s_{\textsl{prepared}}}=H^{\otimes n}\ket{s_{\textsl{initial}}}=\ket{+}^{\otimes n}$
\item One searches for a marked (goal) state $\ket{g}$ out of $N$ eligible states $\ket{g_1},\ket{g_2},\ldots,\ket{g_N}$.
\end{itemize}
{\bf For $1 \le r \le R$ do:}
\begin{itemize}
\item {\it Phase Oracle:} Perform the oracle operation\\ $U_f = \mathds{1} - 2\ket{g}\bra{g}$, 
which maps $\ket{g} \mapsto -\ket{g}$, while leaving all the other vectors unchanged.
\item {\it Diffuser:} Apply the diffuser operator\\
$V = \mathds{1} - 2\ket{s_{\textsl{prepared}}}\bra{s_{\textsl{prepared}}} 
$
\end{itemize}
{\bf Measurement:} Mesure the final state $\ket{s_{\textsl{final}}}$.

{\bf Output:} 
Output the probability vector $(p_1,p_1,\ldots,p_N)$ related to $\ket{s_{\textsl{final}}}$. $p_i$ is an estimation of $\textrm{Prob}(\ket{g}=\ket{g_i})$, $1\le i \le N$.
\end{algorithm}

The actual performance of the algorithm is highly dependent on its circuit implementation, and this even for ideal noiseless models, provided by quantum simulators. 
To make the difference clear, the section below introduces several versions of GSA circuits, along with simulation results. 

\subsection{GSA Circuits and their Simulation}\label{sec:circuits}

A first specificity of quantum circuits is in that they might have or not contain auxiliary (ancilla) qubits. For various GSA implementations for $N=8$, the reader is for instance referred to Fig.~1c of \cite{3qubitGrover-IonQ-Figgatt-2017} for a circuit with one ancilla qubit, to Fig.~4b of \cite{3qubitGrover-Roy-2020} for a circuit without ancilla qubits and to \cite{ElKaderi2022} for circuits with and without ancilla qubits. The results presented in \cite{ElKaderi2022} suggest that free-ancilla qubit circuits perform better than those with ancilla qubits. 

Let us focus on a more involved case of $N=16$. This case is covered by the SDP circuit presented in \cite{Wang2022} and detailed  below. 
Also, in what follows we suggest two new GSA circuits for $N=16$ which are slight extensions of GSA circuits for $N=8$ from \cite{3qubitGrover-IonQ-Figgatt-2017}. 

\subsubsection{SDP circuit \cite{Wang2022}}
\begin{figure*}[t]
  \centering
  \includegraphics[width=\linewidth]{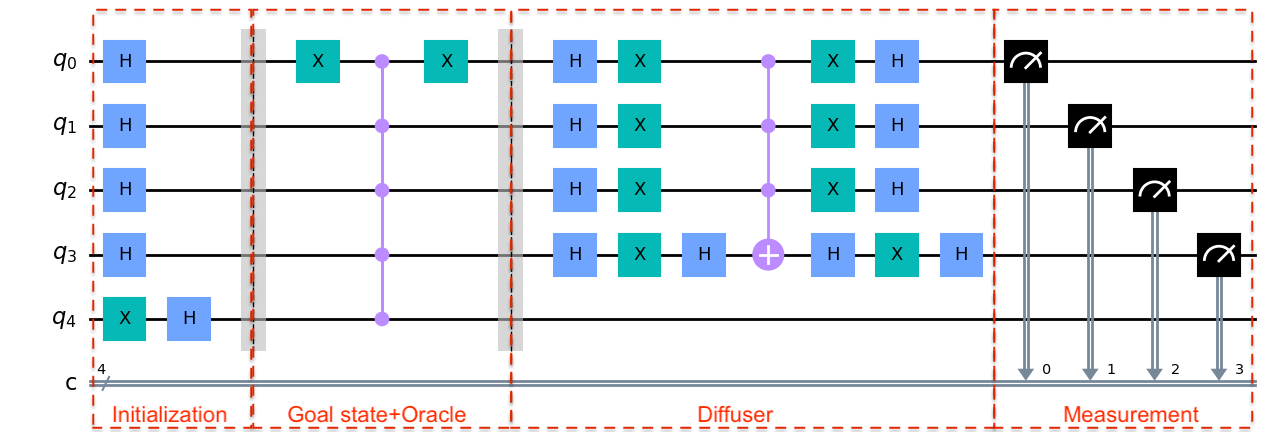}
  \caption{Circuit for the SDP for $N=16$ from \cite{Wang2022} with $\ket{1110}$ as the goal (marked) state and $R=1$. $q_4$ is the ancilla qubit. }
  \label{SDPcir}
\end{figure*}

Fig.~\ref{SDPcir} shows the SDP circuit with 4 data qubits $q_0$, $q_1$, $q_2$, $q_3$, and one ancilla qubit $q_4$. The SDP circuit follows closely the implementation of Algorithm \ref{alg:Grover}.
At the initialization stage, Hadamard (H) gates are applied to data qubits in order to get $\ket{s_{\textsl{prepared}}}$. 
Note that the ancilla qubit is prepared in the $\ket{-}$ state by successive application of $X$ and $H$ gates.
A phase oracle over all qubits is further applied by the MCMT-CZ gate\footnote{For simplicity of presentation the MCMT-CZ gate is not defined here, the reader is invited to refer to \cite{Qiskit}.} from \cite{Qiskit} (drawn in violet in Fig.~\ref{SDPcir}). 
Note that two $X$ gates applied to $q_0$ before and after the MCMT-CZ gate serve to fix $q_0$ to $0$ (and other data qubits will therefore be fixed to $1$).
Thus the goal state in the example of Fig.~\ref{SDPcir} is $\ket{g}=\ket{q_3q_2q_1q_0}=\ket{1110}$. 
The diffuser operator applied after the phase oracle is in fact the phase oracle targeting the $\ket{0000}$ state, put in between H gates. 
Note that the phase oracle and the diffuser operations are to be repeated $R$ times, before proceeding to the final measurement of $\ket{s_{\textsl{final}}}$.
The performance of the SDP circuit on a \textit{qasm} simulator \cite{Qiskit} (i.e. equivalent to a noiseless quantum device) is given in Fig. \ref{Histo_SDP} in blue. 
All the possible states $\ket{g_1},\ldots, \ket{g_N}$ are presented on the $x$-axis, while the $y$-axis gives the value of associated probabilities.
One can see that the goal state $\ket{1110}$ has been correctly detected as it has a higher probability with respect to all other states. 

\begin{figure}[t]
  \centering
  \includegraphics[width=0.7\linewidth]{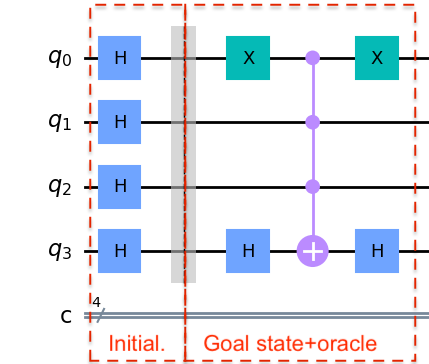}
  \caption{First two stages (\textsl{Initialization} and \textsl{Oracle}) of the GSA circuit for $N=16$ with the goal state $\ket{1110}$, when $R=1$. \textsl{Diffuser} and \textsl{Measurement} stages of the GSA are similar to those in Fig. \ref{SDPcir}. No ancilla qubits are present.}
  \label{GSAcir}
\end{figure}

\subsubsection{GSA with no ancilla qubit}

We suggest a new GSA circuit for $N=16$ with only 4 qubits $q_0$, $q_1$, $q_2$ and $q_3$.
The initialization and the phase oracle stages of our circuit are shown in Fig. \ref{GSAcir}, and the diffuser and measurement stages are similar to those presented in Fig. \ref{SDPcir}. Note that, for the phase oracle, the MCMT-CZ gate used in the SDP circuit, is replaced by the $C^3Z$ gate\footnote{In Fig. \ref{GSAcir}, the $C^3Z$ gate is implemented as a $C^3X$ gate with $H$ gates before and after the gate $X$, by using the fact that $HXH=Z$.}.
The performance of the GSA circuit on a \textsl{qasm} simulator is given in Fig. \ref{Histo_GSA} in blue. 

\subsubsection{GSA with one ancilla qubit}
We suggest another GSA circuit with 4 data qubits $q_0$, $q_1$, $q_2$, $q_3$, and one ancilla qubit $q_4$. This circuit is very similar to the one presented in Fig.~\ref{SDPcir}, with only one exception: the MCMT-CZ gate of the phase oracle is replaced by the $C^4X$  gate. Otherwise, as for SDP, $q_4$ is prepared in $\ket{-}$ as before, and the diffuser and measurement circuits are the same.
Note that the performance of this circuit on \textsl{qasm} is the same as shown by the blue histogram in Fig. \ref{Histo_GSA}.

One can see that the difference between the three circuits above is in the presence/absence of ancilla qubits and in the implementation of the phase oracle. We will see in the next section that this difference is essential for the circuit performance on real quantum devices.    

\section{Performance Evaluation on NISQ Devices}
\label{sec:results}

Let us run the circuits presented in Section \ref{sec:circuits} on real quantum devices, available on an IBM-Q cloud, in particular on \textsl{ibm\_oslo} (7 qubits), \textsl{ibm\_algiers} (27 qubits) and \textsl{ibm\_canberra} (27 qubits). 
It is to note that all the results reported in this section have been obtained by means of Monte-Carlo tests with 1000 experiments (if simulation) or 4000 experiments (if test on real device).

\begin{figure}[t]
     \centering
     \includegraphics[scale=0.62]{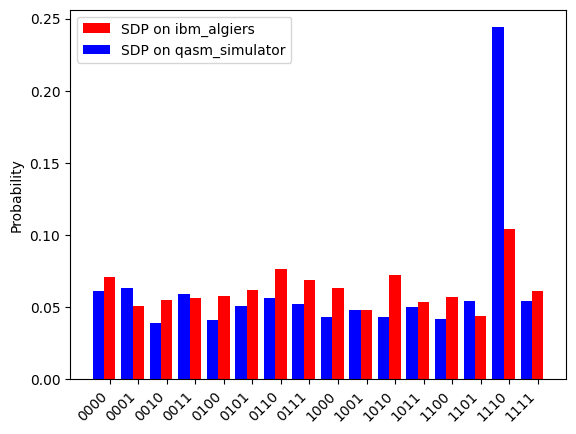}
     \caption{Outcome of SDP \cite{Wang2022} on \textsl{qasm\_simulator} and \textsl{ibm\_algiers}, $R=1$.}
     \label{Histo_SDP}
\end{figure}

\begin{figure}[t]
     \centering
     \includegraphics[scale=0.62]{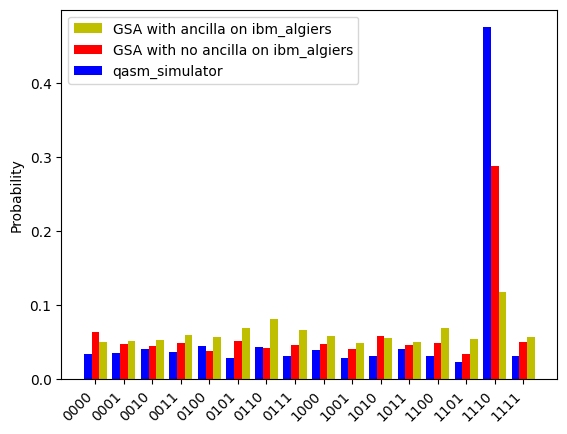}
     \caption{Outcome of GSA with/without ancilla qubit on \textsl{qasm\_simulator} and \textsl{ibm\_algiers}, $R=1$.}
     \label{Histo_GSA}
\end{figure}

\begin{figure}[t]
     \centering
     \includegraphics[scale=0.62]{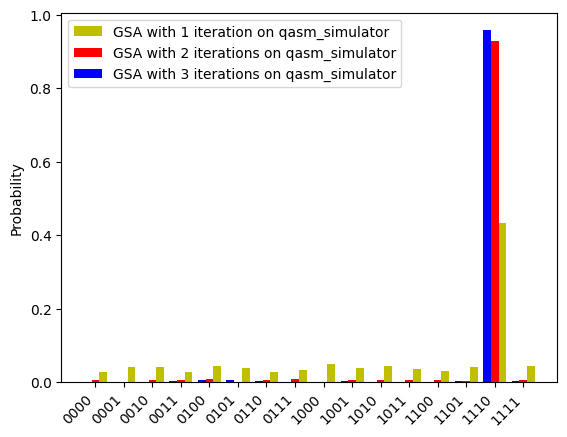}
     \caption{Implementation of GSA on \textsl{qasm\_simulator} for different values of $R$.}
     \label{Hist_Sim}
\end{figure}

\begin{figure}[t]
     \centering
     \includegraphics[scale=0.62]{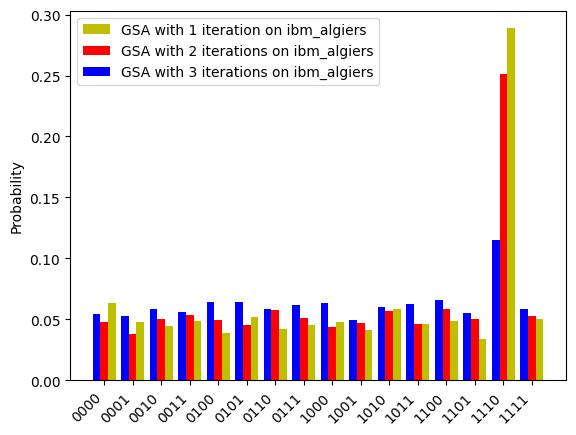}
     \caption{Implementation of GSA \textsl{ibm\_algiers} for different values of $R$.}
     \label{Hist_Qdev}
\end{figure}

\subsubsection{Comparison of outcomes for a simulator and for a NISQ device}
Fig. \ref{Histo_SDP} shows the output of a SDP circuit  in Fig. \ref{SDPcir} \cite{Wang2022} for $\ket{g}=\ket{1110}$, obtained on \textsl{qasm\_simulator} and on the device \textsl{ibm\_algiers}. 
We can observe that the probability associated with the state $\ket{g}$ is larger w.r.t. the probabilities of other states. 
Also, the results on \textsl{ibm\_algiers} are more noisy than the ones from \textsl{qasm}.
A similar observation can be made for Fig. \ref{Histo_GSA}, which compares outputs of GSA with and without ancilla qubit,  for $\ket{g}=\ket{1110}$, tested both on \textsl{qasm\_simulator} and \textsl{ibm\_algiers}. Both GSA circuits behave almost equally on the simulator (blue histogram), but their outcomes are more noisy on \textsl{ibm\_algiers}.
Moreover, the GSA without ancilla qubit is more resistant to the noise induced by the quantum device and has better performance (red histogram). 
Finally, one notices that the GSA with no ancilla qubit is the most efficient circuit among the three circuits introduced in Section \ref{sec:circuits}. 

\subsubsection{Tests for different values of $R$ on a simulator and on a NISQ device} 
Consider the GSA circuit without ancilla qubit. 
Fig. \ref{Hist_Sim} (resp. Fig. \ref{Hist_Qdev}) illustrates the outcome of the circuit on the \textsl{qasm\_simulator} (resp. on \textsl{ibm\_algiers}) for $R=1;2;3$. 
Note that Algorithm \ref{alg:Grover} is shown to converge with the number of repetitions $R$. 
This is exactly what one observes in Fig.~\ref{Hist_Sim} (outcome on the simulator). However, Fig. \ref{Hist_Qdev} shows that the outcome gets worse with $R$.
This is due to the increase of the depth (i.e. the quantity of gates in the circuit) with $R$ and to the accumulation of the noise induced by the gates. 

\begin{table}[t]
\begin{center}
\begin{tabular}{|l|c|c|c|}
\hline
\backslashbox{Circuits}{Devices}
& \textsl{ibm\_algiers}  & \textsl{ibm\_oslo} & \textsl{qasm\_simulator}\\
\hline
SDP    & 4.5   & N/A  & 7.2  \\
GSA with ancilla    & 8.3   & 0.1  & 13.7  \\
GSA w/o ancilla   & 10.8   & 5  & 13.37  \\
\hline
\end{tabular}
\end{center}
\caption{\label{table:n=3} Selectivity for $N=8$ with $\ket{g}=\ket{110}$.}
\end{table}

\begin{table}[t]
\begin{center}
\begin{tabular}{|l|c|c|c|}
\hline
\backslashbox{Circuits}{Devices}
& \textsl{ibm\_algiers}  & \textsl{ibm\_canberra} & \textsl{qasm\_simulator}\\
\hline
SDP    & 1.36   & 1.9  & 6.38  \\
GSA with ancilla    & 4   & 3.25  & 12.8  \\
GSA w/o ancilla   & 6.54   & 4.57  & 9.46  \\
\hline
\end{tabular}
\end{center}
\caption{\label{table:n=4} Selectivity for $N=16$ with $\ket{g}=\ket{1110}$ and $R=1$.}
\end{table}

\subsubsection{Circuits vs. types of devices}
Given the results above, we are ready to make a more complete comparison of circuits run on different devices.

Assume a GSA circuit is run over some NISQ device. Let $g$ be the index of the goal (marked state). Given a probability vector $(p_1,p_2,\ldots, p_N)$ as outcome, define the \emph{selectivity} $S_g$ \cite{YulunKrstic2020} as follows:
\begin{equation}
    S_g = 10 \log_{10} \frac{p_g}{\max_{i \not = g} p_i}.
\end{equation}
In fact, the selectivity plays the same role as signal-to-noise ratio in communication systems. 
It is considered \cite{YulunKrstic2020} that, if $S_g \ge 3$, then the circuit is assumed to recognize the goal state $g$ successfully.

Tables \ref{table:n=3} and \ref{table:n=4} show the values of selectivity for the three GSA circuits from Section \ref{sec:circuits}, implemented on two quantum devices, as well as on the \textsl{qasm\_simulator}, for $N=8$ and $N=16$ respectively. 
Note that the GSA circuit performs the best on quantum devices and the SDP the worst. 
This observation can be explained by the number of gates within the circuit. Indeed, the GSA circuit contains the least number of gates w.r.t. the others, while the SDP circuits contains the most. 
Moreover, the selectivity of a given circuit is the best on \textsl{qasm\_simulator} (perfect running environment) and is worse for quantum devices. It also differs from one device to another due to their differences in implementation (i.e. connectivity of qubits on the circuit) and non-equal calibration parameters. 
The obtained results also warn that the circuit comparison based on simulation results might not reflect the circuit behaviour on real devices. If one only considers the \textsl{qasm\_simulator} column, one would say that the GSA with ancilla qubit is the best circuit. Meanwhile, if one considers results on quantum devices, the GSA without ancilla qubit is the one that distinguishes the goal state for all circuit examples.  

\subsubsection{Extension of GSA to $N=32$}
\begin{figure}[t]
     \centering
     \includegraphics[scale=0.38]{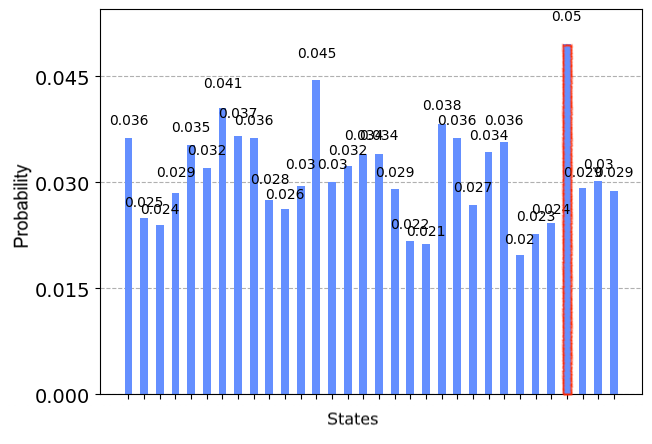}
     \caption{Implementation of GSA without ancilla qubit, for $N=32$ and $\ket{g}=\ket{11100}$, on \textsl{ibm\_algiers}. The bar related to $p_g$ is underlined in red.}
     \label{5-qubit}
\end{figure}

We extended the GSA circuit without ancilla qubit to the case of $N=32$ ($n=5$). The results obtained on a real device (see Fig.~\ref{5-qubit} for tests on \textsl{ibm\_algiers}) are not encouraging: as the number of gates in the circuit grows with $n$, the selectivity $S_g$ takes values close to $0$, meaning that the goal state is indistinguishable from the others. 
This result strongly suggests that, in order to be able to run GSA circuits on real devices for larger values of $N$, 
a noise-reducing method is to be implemented along with the initial algorithm. 

\section{Discussion}
\label{sec:discussion}
The example of the Grover's search algorithm, presented in this work, confirms some implementation rules of quantum algorithms on NISQ devices: if possible, one should avoid a high number of quantum gates in the circuit and/or ancilla qubits.   
In case of GSA, this also means to keep the value of $N$ relatively low. Indeed, for $N=32$ one seems to attain the limit of the GSA circuit feasibility. 

It is possible to improve the performance of the initial algorithm by adapting it to the behaviour of quantum devices, like is has been suggested in \cite{Grover-Zhang-2020} with the aim to improve the GSA through hybrid search algorithm, optimization, and divide-and-conquer search algorithm. However, this is a tedious approach which needs reconsidering every algorithm in particular, and it might not be sufficiently general in practice.

Another approach is the quantum error mitigation (QEM) \cite{Cai2022}, the goal of which is to minimize the mean square error between the perfect circuit state and the noisy circuit state. The QEM is an interesting approach to perform the calibration of the quantum device of the system level, but it can also be incorporated into the circuit level. 

Finally, the third approach is the addition of error-control coding (QEC) to a quantum circuit. 
To show the utility of error-correction schemes for NISQ devices is an open challenging question -- in most of cases the code rate of the quantum error-correction schemes is quite low, namely of order $1/10$, thus to add such a circuit implies an important increase in the number of quantum gates in it, which in its turn increases the induced circuit noise. 
Another alternative would be the use of higher-rate error-detecting codes together with a QEM, like it was first tried in \cite{Pokharel2022}.
In general, a systematic approach connecting QEM methods to QEC and merging them under a common framework of error suppression is another open question.

\section{Acknowledgements}
The authors acknowledge use of the IBM-Q in this work.

\bibliographystyle{IEEEtran}
\bibliography{biblioITW}
\end{document}